\pdfoutput=1

\documentclass[conference,letterpaper]{IEEEtran}
\IEEEoverridecommandlockouts
\usepackage{cite}
\usepackage{amsmath,amssymb,amsfonts}
\usepackage{algorithmic}
\usepackage{graphicx}
\usepackage{textcomp}
\usepackage[T1]{fontenc}
\usepackage[stretch=30,shrink=30]{microtype}
\usepackage{balance}
\usepackage{tikz}
\usepackage{booktabs}
\usepackage{multirow}
\usepackage{makecell}
\usepackage{rotating}
\usepackage{adjustbox}
\usepackage{array}
\usepackage{pifont}
\usepackage{enumitem}

\usepackage[caption=false,font=footnotesize]{subfig}

\usepackage{xurl}
\usepackage[hidelinks]{hyperref}
\hypersetup{
  pdftitle={SigmaTrace: Library-Agnostic Observability for CPU/GPU-Initiated RDMA Communication},
  pdfauthor={Javid Baydamirli, Didem Unat},
  pdfsubject={Accepted author manuscript, SC26, November 15--20, 2026},
  pdfkeywords={SigmaTrace, RDMA, GPU-initiated networking, profiling,
    tracing, observability, libibverbs, mlx5dv, InfiniBand, RoCE, NCCL,
    RCCL, NVSHMEM, High performance computing, Network interfaces,
    Performance evaluation},
  pdflang={en-US},
  pdfdisplaydoctitle=true
}

\DeclareRobustCommand{\circled}[1]{%
  \tikz[baseline=(char.base)]{%
    \node[draw,circle,inner sep=0pt,minimum size=1.0em,
          font=\scriptsize] (char) {#1};}}

\ExplSyntaxOn
\NewDocumentCommand{\code}{m}
{
  \texttt
  {
    \tl_set:Nn \l_tmpa_tl { #1 }
    \regex_replace_all:nnN { ([a-z0-9])([A-Z]) } { \1\c{allowbreak}\2 } \l_tmpa_tl
    \tl_replace_all:Nen \l_tmpa_tl { \char_generate:nn { `_ } { 8 } } { \_\allowbreak }
    \tl_replace_all:Nnn \l_tmpa_tl { @ } { @\allowbreak }
    \tl_replace_all:Nnn \l_tmpa_tl { / } { /\allowbreak }
    \tl_use:N \l_tmpa_tl
  }
}
\ExplSyntaxOff

\usepackage{listings}
\definecolor{codegreen}{rgb}{0,0.6,0}
\definecolor{codegray}{rgb}{0.5,0.5,0.5}
\definecolor{codepurple}{rgb}{0.58,0,0.82}
\definecolor{backcolour}{rgb}{0.95,0.95,0.92}
\definecolor{commentgray}{gray}{0.45}

\lstdefinestyle{cppstyle}{
  language=C++,
  backgroundcolor=\color{backcolour},
  commentstyle=\color{commentgray},
  keywordstyle=\color{magenta},
  numberstyle=\tiny\color{codegray},
  stringstyle=\color{codepurple},
  basicstyle=\ttfamily\footnotesize,
  breakatwhitespace=false,
  breaklines=true,
  captionpos=b,
  keepspaces=true,
  numbers=left,
  numbersep=5pt,
  showspaces=false,
  showstringspaces=false,
  showtabs=false,
  tabsize=2,
}
\begin{document}

\title{SigmaTrace: Library-Agnostic Observability for
CPU/GPU-Initiated RDMA Communication%
\thanks{Accepted for publication in the International Conference for High
Performance Computing, Networking, Storage and Analysis (SC26),
November 15--20, 2026, Chicago, Illinois, USA. Author's accepted manuscript.
\copyright\ 2026 IEEE. Personal use of this material is permitted.
Permission from IEEE must be obtained for all other uses, in any current
or future media, including reprinting/republishing this material for
advertising or promotional purposes, creating new collective works,
for resale or redistribution to servers or lists, or reuse of any
copyrighted component of this work in other works.}}

\author{\IEEEauthorblockN{Javid Baydamirli}
  \IEEEauthorblockA{
    \textit{Computer Science and Engineering, Koç University} \\
    Istanbul, Turkey \\
  jbaydamirli21@ku.edu.tr}
  \and
  \IEEEauthorblockN{Didem Unat}
  \IEEEauthorblockA{
    \textit{Computer Science and Engineering, Koç University} \\
    Istanbul, Turkey \\
  dunat@ku.edu.tr}
}

\maketitle

\newcommand{\ToolName}{\textnormal{\textsc{SigmaTrace}}}

\begin{abstract}
  RDMA networking enables low-la\-tency, high-band\-width communication on large systems.
  Despite its adoption, however, observability is lacking. Existing tools target
  specific communication libraries or only report aggregated data, missing per-operation behavior.
  The gap is widest for GPU-initiated networking, whose in-kernel
  RDMA operations are invisible to the host.

  We present \ToolName{}, a library-agnostic RDMA profiling framework
  providing fine-grained, per-operation visibility for CPU and
  GPU-initiated networking. It traces operations at the
  \texttt{libibverbs}/\texttt{mlx5dv} layers,
  the common denominator beneath communication libraries
  on InfiniBand and RoCE, and employs a low-overhead memory interposition technique
  that, to our knowledge, gives the first observability into
  GPU-initiated RDMA without source patching or binary instrumentation.

  Host-side tracing adds $1.3\%$ overhead on NCCL AllReduce at 128 GPUs
  and under $2\%$ on RCCL at 1024 GPUs.
  On a persistent-kernel BFS and Mixture-of-Experts inference,
  \ToolName{} exposes an in-kernel barrier spread and
  $2.13\times$ inter-node traffic skew invisible to existing profilers.
\end{abstract}


\section{Introduction}
\label{sec:introduction}

Communication overhead is a major consideration for scalability and
efficiency in AI training and inference, as well as scientific computing
workloads \cite{megatron-LM, calculon, padal2}.
To better meet performance-critical data movement demands of modern workloads,
large-scale clusters now ubiquitously employ high-speed and low-latency 
Remote Direct Memory Access (RDMA) networking.
At the software level, RDMA communication has seen wide adoption
in libraries including NCCL \cite{NVIDIA-NCCL-Docs}, NVSHMEM \cite{Nvshmem_Docs},
MPI, UCX \cite{UCX}, and many of their derivatives, dominating the high-performance
computing landscape.

Significant effort has also been directed at bringing native
networking to GPUs. Traditionally,
CPU threads managed network buffers and operations on behalf of GPUs.
GPUDirect RDMA \cite{cuda-aware-mpi}
moved the \emph{data} path onto devices, allowing NICs to read
directly from GPU memory.
Later, GPUDirect Async (IBGDA) \cite{GPUDirect-Async} moved the
\emph{control} path onto devices as well, enabling
GPU threads to construct RDMA messages in device memory, write to NIC
doorbells to initiate communication, and poll for completion without
CPU involvement.
This paradigm of putting the devices in control of communication
can reduce end-to-end latency greatly while allowing higher levels
of computation and communication overlap
\cite{cpu-free, preconditioned-cg, atos, gromacs-halo, aCG}.
GPU-initiated networking has been adopted by first-party libraries
\cite{Nvshmem_Docs, NVIDIA-NCCL-Docs}
and is increasingly specialized for latency-sensitive workloads
such as Mixture-of-Experts (MoE) routing: DeepEP~V2 \cite{deepep2025}
uses NCCL's device-side API to minimize all-to-all dispatch latency
across hundreds of expert-parallel ranks, and several communication
libraries have since built on it \cite{hybridep2026, ncclep2026, uccleep2025}.

Despite the growing adoption of both CPU and GPU-initiated RDMA, tooling
for observability, performance engineering, and communication
debugging has not kept pace. Existing profiling and tracing tools largely
focus on a single high-level layer in the communication stack, missing
RDMA-level information, and remaining entirely blind to
GPU-initiated operations: GPU-initiated RDMA is opaque to
the host and consists of loads and stores that are indistinguishable
from the many other
memory operations occurring within a kernel. Unlike its CPU-side
counterpart, there are
no API boundaries or software events that can be used as observability points.

Tools like Mycroft \cite{Mycroft}, NCCL Inspector
\cite{NVIDIA-Blog-NCCL-Inspector}, and NPKit
\cite{NPKit} that instrument GPU communication libraries
trace the host-side proxy threads and API calls but stop squarely at the
kernel launch. On the other hand, binary instrumentation-based tools
like Snoopie \cite{Snoopie} incur high overheads due to  assembly level
injection across all memory operations \cite{NVIDIA-NVbit}.
Vendor profilers like NVIDIA's Nsight Systems
\cite{NVIDIA-Nsight-Systems} and AMD's ROCm Systems Profiler
\cite{AMD-rocprofsys}
provide system-level aggregate NIC counters, but cannot attribute metrics to
an individual
kernel, GPU, or even process, and offer little visibility into the
actual RDMA operations
being performed. As a result, the most performance-critical
operations are paradoxically
the least visible.

To address the observability gaps, we present \ToolName{}, an
RDMA profiling framework that provides fine-grained,
per-operation visibility into the full communication stack for both CPU
and GPU-initiated networking. Our approach relies on one
observation: while GPU-initiated RDMA bypasses traditional
software interception points such as API calls, it cannot
bypass the hardware interface between the GPU and the NIC.
Even with fully GPU-native networking, where all communication buffers
reside entirely within device memory and data moves through hardware-level
DMA without the CPU on the path, every RDMA operation is ultimately
initiated by a PCIe write. The write is done from within GPU kernels
to a NIC's doorbell register, which is exposed in its PCIe
memory-mapped I/O (MMIO) Base Address Register (BAR) and memory-mapped
into the application's address space. This gives us a single observable
point below all communication libraries and opaque memory
instructions, on a data path accessible to the CPU.

We exploit this point and selectively replace the doorbell pages that
would be mapped into GPU address space---NIC \emph{User Access Region}
(UAR) pages---with pinned CPU memory
through hijacks at the \code{mlx5} direct verbs and CUDA driver levels,
redirecting doorbell writes without modifying the
target application. When the GPU rings the doorbell to
initiate a transfer, our polling thread captures it
first, records the RDMA metadata associated with it,
and forwards the request to the real NIC for execution.
To our knowledge, \ToolName{} is the first tool to both enable
observability into GPU-initiated RDMA operations, and to employ
GPU-level memory interposition for profiling without binary
instrumentation.

Our contributions are as follows:
\begin{itemize}
  \item \textbf{The first observability solution for true
    GPU-initiated networking}:
    We capture precise doorbell timing and RDMA work queue content,
    invisible to all existing tools.
  \item \textbf{UAR memory interposition technique}:
    We detail a novel \emph{Memory Hijack-and-Repeat} mechanism that
    records and redirects GPU memory
    operations without source code modifications, at a fraction of
    the overhead of binary instrumentation.
  \item \textbf{Library-agnostic multi-layer tracing}:
    We target the lowest common denominator layer above hardware---%
    \texttt{libibverbs}
    and \texttt{mlx5dv} \cite{rdma-core}---that upstream libraries
    use on InfiniBand and RoCE platforms. We additionally instrument
    \code{UCX} and \code{libfabric}
    directly, extending support to HPE Slingshot and AWS EFA fabrics.

  \item \textbf{Scalable cross-node correlation and analysis}:
    We develop a minimal-overhead event recording framework with
    hardware-ID-based distributed trace correlation,
    and support both timeline visualization of network events and
    quantitative analysis via Perfetto and PerfettoSQL.
\end{itemize}

We evaluate \ToolName{} on up to 128 nodes, demonstrating as little
as $1.3\%$ overhead on NCCL AllReduce on 128 GPUs across 32 nodes
and under $2\%$ on RCCL at 1024 GPUs across 128 nodes.
We further characterize the overhead of the memory-hijack mechanism
with microbenchmarks and present two case studies:
a distributed persistent-kernel BFS traversal, where \ToolName{}
reveals early per-level graph load imbalance and straggler barriers,
and a Mixture-of-Experts inference workload, where
\ToolName{} exposes inter-node traffic skew across data-parallel ranks.

\section{Background and Motivation}
\label{sec:background}

RDMA is a low-latency and high-throughput data movement technique
that enables direct memory access between peers without OS
intervention. In networking, this allows NICs to read from and write to
system memory directly, moving data over the network with minimal
CPU coordination.

\begin{figure}[htbp]
  \centering

  \includegraphics[width=\columnwidth]{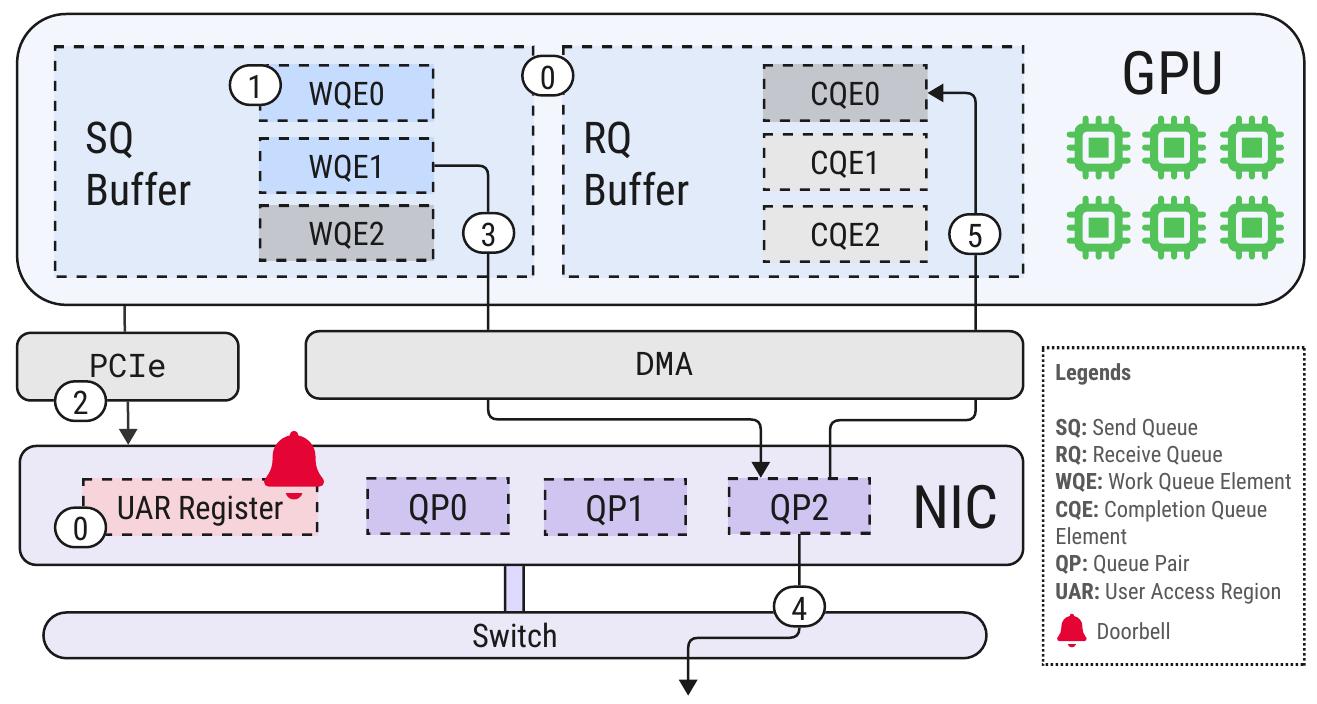}
  \caption{Illustrating the GPU RDMA communication path. \circled{0} The
    communication runtime allocates SQs and RQs in device memory and
    maps the doorbell registers into GPU-accessible address space.
    During communication, \circled{1} GPU threads construct WQEs in
    device memory and \circled{2} ring the doorbells.
    Data flows (\circled{3} and \circled{4}) through DMA engines
    in GPUs and NICs, while doorbells go through PCIe.
    \circled{5} The NIC later posts a CQE to signal completion.}
  \label{fig:motivation-nvshmem}
\end{figure}

RDMA communication is organized around sets of queues. In the
InfiniBand verbs model, endpoints are \textit{Queue Pairs (QPs)},
each consisting of a
\textit{Send Queue (SQ)} and a \textit{Receive Queue (RQ)}. To issue
an RDMA operation,
the application constructs a \textit{Work Queue Element (WQE)} in SQ
memory describing the
operation type (\texttt{WRITE}, \texttt{READ}), local and remote
buffer identifiers, and the transfer size.
The application then rings the NIC \textit{doorbell} through a write
to its \textit{User Access Region (UAR)}, a register page exposed in
the NIC's PCIe BAR and memory-mapped into the process address
space. A doorbell write may
batch multiple WQEs for
asynchronous execution, after which the NIC posts \textit{Completion
Queue Entries (CQE)} to
the \textit{Completion Queue (CQ)}. Data movement other than the
doorbell takes place through
DMA engines within the NIC and PCIe subsystem.

\subsection{GPUDirect RDMA}
\label{sec:background-gpudirect-rdma}
Hardware prerequisites for RDMA networking---PCIe BAR access and DMA
capability---are present on modern GPUs,
which are themselves both DMA peers and PCIe endpoints. This means
that all RDMA queues and memory buffers
described above can reside in GPU memory and be accessed by NIC
hardware transparently.
Furthermore, the NIC's doorbell registers, exposed on its PCIe BAR,
can be mapped into user-space
virtual memory and subsequently registered with the GPU driver for
in-kernel access, allowing
GPU threads to construct WQEs in device memory and notify the NIC via
stores to the mapped doorbell address. This mechanism underlies
GPUDirect RDMA \cite{cuda-aware-mpi}
and GPUDirect Async \cite{GPUDirect-Async}, which enable
GPU-initiated networking (GIN)
\cite{gpuinitiatednetworkingnccl}.

This communication path is used by NVSHMEM's IBGDA
transport \cite{Nvshmem_Docs} and the DOCA GPUNetIO SDK
\cite{NVIDIA-DOCA-GPUNETIO}, used by NCCL's recently introduced
\code{GDAKI} backend \cite{gpuinitiatednetworkingnccl}.
On AMD systems, rocSHMEM \cite{AMD-rocSHMEM-Docs} implements an
analogous GPU-initiated path using GDA
transports over Mellanox NICs, while RCCL \cite{AMD-RCCL-Docs}
currently relies on CPU proxy threads for
network operations.

\subsection{Motivation and Gaps in Observability}
\label{sec:background-motivation}
Consider an application performing a remote \code{put} operation
(Figure~\ref{fig:motivation-nvshmem}).
Internally, GPU threads construct several RDMA \code{WRITE} WQEs of
varying sizes
on their local memory to addresses on remote peers and write to NIC
doorbell registers
using global memory operations (\code{STG.E.STRONG.SYS} or
\code{STG.E.STRONG.GPU} in CUDA SASS). Depending
on the size of the
transfer, the number of peers, and the network topology, a single
kernel invocation may issue hundreds of such operations across several
QPs connected to many remote nodes and GPUs.

Because this communication path is reduced to ordinary memory
operations on mapped address ranges,
GPU-initiated RDMA is largely invisible to conventional profiling
tools. Nsight Systems shows such
kernels as opaque timeline blocks and reports only aggregate NIC
statistics without per-kernel or
per-GPU attribution. Similarly, tools such as Mycroft \cite{Mycroft},
NPKit \cite{NPKit},
ucTrace \cite{ucTrace}, and Aegis \cite{Aegis} instrument host-side
communication and therefore
miss in-kernel operations that bypass the host entirely.

At this stage, extracting runtime network information from the
application is impossible without extensive source code
modifications or binary patching. Performance pathologies
caused by communication imbalance across peers
and QPs, congestion reflected in doorbell register activity,
bandwidth underutilization across kernels, and other correctness bugs
are reflected in doorbell register writes and WQE contents that no existing tool
currently targets.

\section{System Design}
\label{sec:system-design}

\subsection{Design Goals and Challenges}
\label{sec:design-goals}

We define and outline the following design goals for \ToolName{} and
the challenges they introduce:

\begin{itemize}
  \item \textbf{Capturing all RDMA operations}, both CPU and GPU-initiated.
    The tool must record precise doorbell timings, extract WQE
    metadata, and correlate
    events across threads, processes, and network nodes.

  \item \textbf{Maintaining low overhead} while remaining non-intrusive
    to the communication flow.
    Since RDMA applications often issue and poll for fine-grained
    operations, the tool must handle high-frequency traces
    efficiently, scale reasonably alongside the application,
    and consolidate data efficiently on large systems.

  \item \textbf{Remaining communication library agnostic} without
    requiring source modifications or recompilation.
    Since communication routines may traverse multiple software
    layers and internal buffers, the tool must
    gather sufficient information independent of library specifics.

\end{itemize}

\subsection{Architecture Overview}
\label{sec:architecture-overview}
\ToolName{} is implemented as a two-part architecture that decouples
runtime capture and offline reconstruction. The \emph{Profiler} on
the hot path captures fine-grained RDMA activity while keeping the
overhead low, and the \emph{Correlator} parses and consolidates the
collected data post-mortem.

The \textit{Profiler} intercepts communication activity at selected
software layers,
including host-side RDMA and middleware calls, \texttt{mlx5dv} state creation,
and CUDA driver calls required for GPU doorbell interposition. We
keep the \textit{Profiler} as light as possible with minimal state
tracking and only record data points available to a single thread during
application execution.

Subsequently, the \textit{Correlator} merges per-thread traces and
metadata, normalizes
timestamps, reconstructs queue-pair connectivity, associates
doorbells with WQEs,
and matches communication activity across processes and nodes.
This split allows us to
minimize the overhead of the tool, as \textit{Profiler} threads
perform the minimal amount of work to record traces without
synchronizing or communicating among one another.

\subsection{Selective Tracing}
\label{sec:design-selective-tracing}
The \textit{Profiler} is composed of a set of modular event providers
implemented as
\code{LD_PRELOAD} shims that can be enabled selectively at runtime.
This allows users to pick and choose injection targets and reduce the
overhead of the tool; for example, a user may choose to omit the
\code{libibverbs} layer and only collect \code{UCX} data at startup
without further runtime overhead.

Within the selected libraries, \ToolName{} supports three trace modes:
\begin{itemize}
  \item \textbf{Aggregate}:
    Collects only total communication volume with source and
    destination markers, using thread-local in-memory aggregation
    with no timestamping for minimal overhead and file I/O.
  \item \textbf{Simple (Default)}:
    Records a reduced set of traces, omitting high-frequency
    operations such as \code{ibv_poll_cq} completions. This mode
    allows a higher-level
    visualization of the network flow at low overhead.
  \item \textbf{Full}:
    Records all supported events, including per-request operations
    such as nested \code{ibv_post_send} work requests. This mode
    provides the highest fidelity at the
    cost of increased I/O pressure and a larger hit to application
    performance, and is meant for short, targeted debugging tasks
    rather than production runs.
\end{itemize}

\begin{figure*}[tbp]
  \centering
  \includegraphics[width=\textwidth]{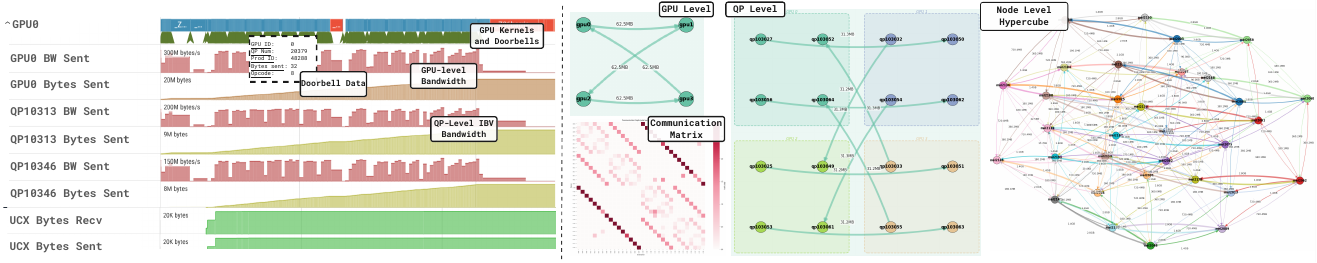}
  \caption{Sample visualized data. Top: Multi-level timeline trace with
    doorbell information, produced by the Simple and Full modes.
    Bottom: Aggregate-mode data at multiple
    granularities (IBV Queue Pairs, GPUs, Ranks, Nodes).
  }
  \label{fig:tool-screenshots}
\end{figure*}

We note that the architecture of \ToolName{} fully disables trace
collection for omitted events,
meaning no overhead is incurred for calls that are not traced.

\section{Host-Side Instrumentation and Trace Collection}
\label{sec:host-instrumentation}
\subsection{Host-side Interception}
\label{sec:host-interception}

In order to intercept and record the doorbells alongside other RDMA
operations, we need to intercept the
host-side \code{mlx5dv} library and a set of CUDA driver APIs.
\ToolName{} primarily uses a function symbol interposition technique
to trace communication
calls and collect data through injected callbacks. This is achieved through the
\code{LD_PRELOAD} mechanism of the dynamic linker on ELF-based systems
which allows us to selectively route library calls to our custom callbacks.

We identify three ways in which communication calls are resolved and
invoked in the target libraries:
(1) Plain link-time binding to shared objects used for bootstrapping
functions like \code{ibv_open_device},
(2) Runtime resolution through \code{dlopen},  \code{dlsym} and \code{dlvsym}
for most library calls, and
(3) Symbol tables on internal library objects for high-frequency calls like
\code{libibverbs@ibv_post_send} and \code{libfabric@fi_writemsg}.

\ToolName{} supports instrumenting all three through simple symbol
shims, injections at
\code{dlsym} and \code{dlvsym}, and efficient object
tracking for high-frequency dispatch table calls.

\subsection{CUDA Driver Interception}
\label{sec:cuda-interception}

To transparently interpose UAR memory without changing pointer
semantics and consequently distorting application execution, \ToolName{}
must modify the behavior of the CUDA driver API, in particular, host
memory management functions. While CUDA provides tooling for
observability into the driver API through CUPTI
\cite{NVIDIA-CUPTI-Docs} and the Sanitizer API \cite{NVIDIA-Sanitizer-API-Docs},
it is currently not possible to inject custom code during execution
or alter return values of the runtime methods. Moreover,
the \code{LD_PRELOAD} and \code{dlsym} interposition techniques
described earlier are not sufficient for the CUDA driver API,
since most symbols are resolved through internal undocumented dispatch tables.

To address this, we intercept two CUDA driver entry points used to
bootstrap symbol resolution: \code{cuGetProcAddress} and
\code{cuGetExportTable}. These interfaces expose dispatch tables that
are later used to resolve additional CUDA driver and runtime
symbols. We instrument the tables by inspecting and
saving the original function pointers and replacing them with
pointers to our own wrappers. This allows us to selectively
inject our instrumentation while preserving the original CUDA
driver behavior for other calls.

\subsection{Doorbell State Reconstruction}%
\label{sec:doorbell-state-reconstruction}

To interpret doorbell writes, \ToolName{} must know the QP
configuration, SQ memory layout, and endpoint information associated
with each doorbell at the time it is fired. We extract this data
by intercepting \code{mlx5dv} setup calls on the host side during
transport initialization.
Specifically, we track the following runtime state:

\textbf{UAR lifecycle.} We record UAR allocations done through
\code{mlx5dv_devx_alloc_uar} with their register address, page
identifier and the associated NIC device context. On a call to
\code{mlx5dv_devx_free_uar} with the same page ID and register
address, the tracking state is cleared.

\textbf{UMEM registration.} \code{mlx5dv_devx_umem_reg} registers a
contiguous memory region (possibly GPU memory allocated with
\code{cuMemAlloc} or equivalent) into the NIC's address space. \ToolName{} queries
\code{cudaPointerGetAttributes} on the registered address to
determine whether the memory is GPU-resident, or host-accessible. For
GPU-resident memory, we record a mapping from the registered-memory
ID to the memory address, GPU device ID, memory type, and CUDA
context. This mapping allows us to determine which GPU owns the send
queue buffer when a queue pair referencing
the memory ID is created.

\textbf{QP creation}. Following the UAR allocation and UMEM
registration, a queue pair is created through
\code{mlx5dv_devx_obj_create} with opcode
\code{MLX5_CMD_OP_CREATE_QP}, referencing the aforementioned UAR and UMEM.
We extract  the address, capacity, and byte offsets of the SQ, the
service type of the transport, and the
QP number from the command mailbox.

\subsection{Post-Mortem Analysis and Correlation}
\label{sec:post-mortem-analysis}
After execution, the \textit{Correlator} reads the raw event files in
parallel and merges them into a unified event stream.
It performs the following passes to reconstruct communication
activity across threads,
processes, and nodes:

\begin{figure}[htbp]
  \centering
  \includegraphics[width=\columnwidth]{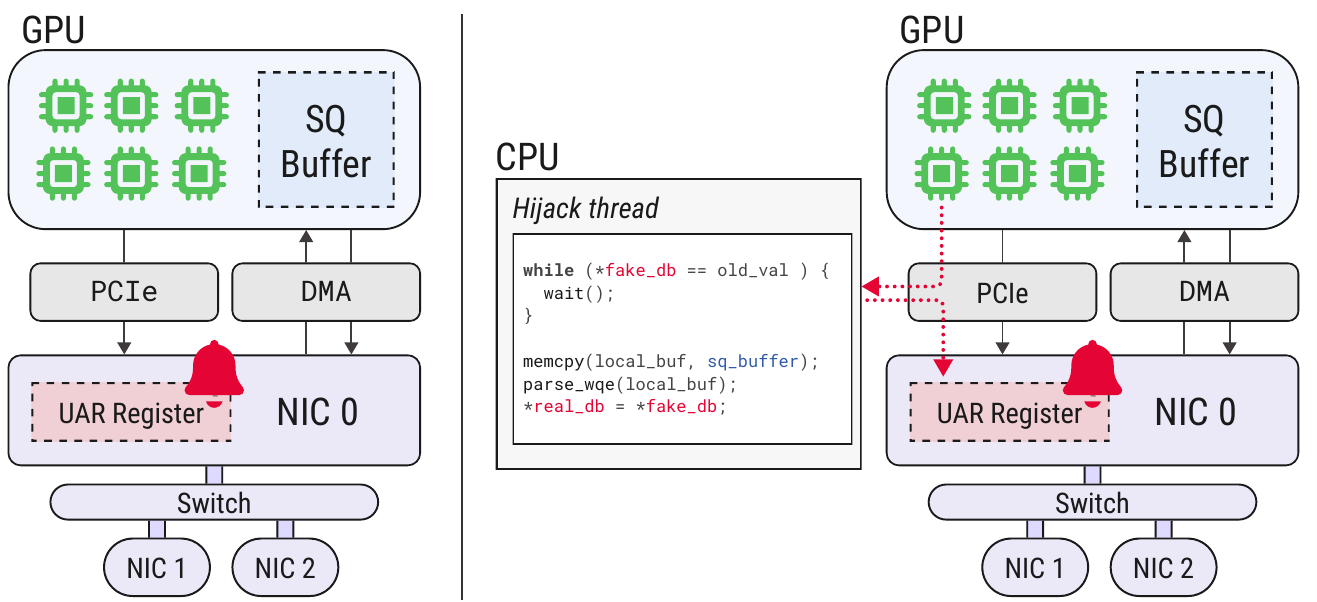}
  \caption{High-level overview of the doorbell hijack mechanism.
    \textbf{Left}: regular GPU-initiated communication flow.
    \textbf{Right}: our hijack thread transparently intercepting
    doorbell writes, reading from the SQ and forwarding to the NIC.
    No difference in execution or semantics from the perspective of
    the GPU or NIC.
  }
  \label{fig_doorbell}
\end{figure}

\textbf{Timestamp normalization}.
Per-thread timestamps are recorded using the x86 timestamp counter
(\texttt{RDTSC}), which
provides cycle-accurate timing with negligible overhead.
To establish a consistent time base and to avoid inter-node clock
drift during distributed trace correlation, the \textit{Profiler}
records pairs of wall clocks synchronized via the Precision Time
Protocol (PTP) at initialization. The
\textit{Correlator} converts per-thread cycle counts to
nanoseconds using this calibration data.
This allows events collected from different threads and processes
to be compared consistently on a unified timeline.

\textbf{QP topology reconstruction}. Using the metadata collected
during QP creation and state transitions, the \textit{Correlator}
reconstructs the connectivity between local and remote QPs. This
topology information is used to determine the communication endpoints
associated with each event.

\textbf{Doorbell-to-WQE association}. Doorbell events are matched
with the corresponding WQE using QP identifiers and producer state.
This step annotates each doorbell with communication metadata such as
operation type, remote target address, and transfer size.

\textbf{Export formats.}
The correlated trace is exported in Perfetto Protobuf format for
timeline visualization with flow arrows connecting send/receive
across nodes and custom tracks for doorbell events alongside kernel
timelines. Additionally, the trace can be queried with PerfettoSQL
for custom filtering, aggregation, and post-mortem analysis.

\textbf{Visualization.} The collected data can be presented in two
forms: as a unified timeline trace, or as aggregate data visualized
in communication matrices and graphs.
Figure~\ref{fig:tool-screenshots} shows sample outputs
from various configurations.
The trace data is exported with normalized timestamps and custom
tracks for GPU-level doorbell activity. Aggregate data,
including per-QP bandwidth, per-GPU communication volume, and
node-level communication matrices, is stored in JSON format
and can be visualized through standard tools or queried directly for
programmatic analysis.

\section{GPU Doorbell Hijack and WQE Recovery}
\label{sec:doorbell-hijack}

\subsection{Memory Hijack}
\label{sec:memory-hijack}

Our doorbell tracking mechanism involves
selectively replacing memory-mapped NIC UAR pages with
CPU-visible memory pages to intercept doorbell writes without source
code modifications or binary instrumentation.
To achieve this, we identify and exploit the following three-step
sequence for setting up GPU-initiated networking:

\begin{enumerate}
  \item \textbf{UAR Allocation}: The target application calls
    \code{mlx5dv_devx_alloc_uar}
    to allocate a NIC user access region. The allocated structure
    contains a host-accessible
    virtual address space pointer to the doorbell register mapped
    over the NIC's PCIe BAR.
  \item \textbf{CUDA Host Memory Registration}: The target
    application then calls
    \code{cuMemHostRegister} on the register address to make the
    mapped I/O range accessible
    to CUDA kernels.
  \item \textbf{Device Pointer Resolution}: The application calls
    \code{cuMemHostGetDevicePointer}
    to obtain a GPU-side pointer to the register address. This
    pointer is written to by GPU threads
    to ring doorbells.
\end{enumerate}

We intercept all three steps above to transparently replace the UAR
register space as well as device-side
pointers. At steps 1 and 2, we simply record and mark the address
ranges as possible candidates for replacement.
At step 3, we allocate a \emph{fake} 4KB page in pinned host memory
and return a device-side pointer to it instead.
From the GPU's perspective, stores to pinned host memory and to the
registered UAR mapping behave identically---both are uncached
posted writes that cross PCIe---%
so the profiled application transparently receives a valid device-side
pointer with the same store semantics as the original address.

\subsection{Memory Repeater and Polling}
\label{sec:memory-repeater}
A dedicated CPU thread is created for each unique doorbell to monitor
the first 64
bytes of its associated replacement page for changes. When a write is
detected, the thread
invokes a callback for event logging, and forwards the written value
to the real NIC UAR. Figure~\ref{fig_doorbell}
illustrates this mechanism.

To keep CPU utilization low without increasing wake-up latency,
we use \texttt{monitorx}/\allowbreak\texttt{mwaitx} (AMD) or
\texttt{umonitor}/\allowbreak\texttt{umwait} (Intel) for polling. These
instructions block the thread until the cache line accessed
by the GPU is invalidated. GPU doorbell writes traverse PCIe and are
delivered as posted writes to the CPU's coherence domain,
invalidating the monitored cache line and waking the polling thread.
Since \code{mlx5} doorbell writes are exactly
8 bytes, each write invalidates a single cache line and wakes the
polling thread, and no spurious wake-ups occur from unrelated memory
traffic, as each \textit{fake} page is dedicated to one doorbell and
shared with no other data structure. Back-to-back doorbells that land
before the polling thread wakes may coalesce into a single
observation; this loses no information, because doorbell values carry
a monotonically increasing producer index, so the latest observed
value covers all preceding writes and WQE recovery
(\S\ref{sec:doorbell-wqe-parsing}) accounts for the full
producer-index delta.

Parked polling threads do not contend with the application for CPU
resources: while waiting in \code{mwaitx}/\code{umwait}, a core
consumes no execution bandwidth and yields its resources to the
sibling hyperthread. Since QPs have no core affinity, polling threads
can additionally be pinned to any core, away from application threads.

\subsection{Doorbell and WQE Parsing}
\label{sec:doorbell-wqe-parsing}

Each \code{mlx5} doorbell write encodes the target QP number and a
producer index that identifies the position of
the most recent WQE in the send queue.
We extract these fields to associate doorbell events with the communication
requests issued by the GPU.

While the doorbell control sequence is written to (supposed) NIC
memory that is readily accessible to CPU threads after interposition,
the associated work queue elements and the data ranges they describe
can reside in pure GPU memory. As such, individual
WQEs are not directly readable on the CPU side at the time of the
doorbell ring and
require a device-to-host copy. Since doorbells happen frequently on
latency-critical
paths, we cannot synchronously copy and parse these structures on
every doorbell write
without a high performance penalty. Instead,
we adopt a strategy where we split WQE recovery
into a cheap per-doorbell step and a deferred batched parse step.

On every doorbell, the polling thread records the timestamp and
producer index, enqueues an \emph{asynchronous} device-to-host copy
of the newly posted WQE slots on a dedicated CUDA stream, and
forwards the doorbell to the NIC. Because the copy does not need
to be synchronous, forwarding is not delayed further.

Parsing the recovered WQEs is deferred and done in batches: once a number of
new WQE slots have accumulated since the last batch, the polling thread
synchronizes the copy stream and parses the recovered entries before
forwarding the current doorbell. These threshold-crossing doorbells
are blocked on copy completion, while most doorbell writes are forwarded without
deferral. This amortizes the synchronization and parsing costs across many
doorbell writes while preserving accurate per-doorbell timing.

This design is safe with respect to send queue wrap-around.
Since the GPU can only overwrite a slot after the circular buffer
wraps to avoid data corruption, batching at less than or equal to the number
of available slots keeps us ahead of the wrap-around.
We found a batch threshold of 128 to be safe for the workloads we tested,
though, depending on the communication library configuration, it can be increased
further. NVSHMEM's IBGDA transport allocates 1024 WQE slots per send queue
(\code{log_wq_size} = 10) where a batch size of 512 can be used, and
DeepEP similarly sizes its queues larger than its maximum number of
in-flight requests to skip slot-availability checks. 

We note that corruption in WQE recovery is detectable by inspecting
the monotonic producer index. Alternatively, batching can be tightened or
disabled entirely to copy and parse every doorbell and its associated WQEs
synchronously at the expense of a higher overhead.

After each batch copy, we parse the relevant WQE control
segments to recover the operation type (\texttt{opmod}),
the number of data segments (\texttt{ds}), and the data
segments themselves, which contain the remote address
(\texttt{raddr}), remote memory key (\texttt{rkey}), and
the byte count for each scatter/gather entry.

By correlating deferred WQE metadata with previously recorded
doorbell timestamps, \ToolName{} reconstructs fine-grained RDMA
communication behavior: per-operation timing, per-QP traffic
volume, and per-peer communication patterns for GPU kernels
that are otherwise completely opaque to existing tools.

\textbf{Memory ordering considerations.}
A potential concern is whether WQE contents associated with doorbells are
visible to the CPU when we copy them back. The correctness of GPU-initiated RDMA itself
requires that all WQE stores reach the GPU's coherence point before the
doorbell write reaches the NIC. The NIC requires this ordering for its
own DMA read of the WQEs following doorbell rings. Communication libraries
therefore order WQE stores ahead of the doorbell store with a memory fence or a release store
to ensure correct delivery; NVSHMEM, for example, issues a \code{__threadfence} 
or release-store (\code{st.release}) sequence when ringing the doorbell.
This ordering is established on the GPU before the doorbell store is issued, and
is thus independent of where that store lands. It holds for our replacement page
identically as it does for the real UAR. The WQEs are read back after the doorbell
is observed by the copy engine, which observes GPU memory at the same scope as the NIC's DMA read. Once the polling thread observes a new doorbell value, every WQE
slot covered by its producer index is already visible to our device-to-host copy,
as it would have been to the NIC.

\section{Evaluation and Case Studies}
\label{sec:evaluation}

We evaluate \ToolName{} in two parts. First, we quantify its
overhead in two experiments: the per-operation cost of GPU-initiated
doorbell interception (\S\ref{sec:evaluation-doorbell}) and
end-to-end application overhead at multi-node scale
(\S\ref{sec:evaluation-scaling}).
We then demonstrate the observability gains on two real workloads:
distributed BFS (\S\ref{sec:evaluation-bfs})
and Mixture-of-Experts inference
(\S\ref{sec:evaluation-moe}).

\subsection{Experiment Setup}
\label{sec:evaluation-setup}
Each experiment is repeated five times on a fixed set of compute
nodes within an allocation.

\paragraph{Single-node IBGDA platform}
Doorbell interception experiments run on a single node with four
NVIDIA A100-SXM4-40GB GPUs connected to one Mellanox ConnectX-5
(MT27800) InfiniBand Host Channel Adapter (HCA) via PCIe Gen4.
We disable NVLink peer-to-peer transfers\footnote{\code{NVSHMEM_DISABLE_P2P=1}}
and force all inter-GPU
communication through the NIC using NVSHMEM's GPU-initiated IBGDA
transport\footnote{\code{NVSHMEM_REMOTE_TRANSPORT=ibdevx}}\textsuperscript{,}\footnote{\code{NVSHMEM_IBGDA_NIC_HANDLER=gpu}}.
This configuration routes every GPU-initiated doorbell write through
a single HCA, producing the maximum possible doorbell rate per NIC
and is a worst-case stress test for UAR interposition.
Normally, systems with per-GPU NICs
distribute doorbell traffic across HCAs, reducing per-NIC
interposition load proportionally. The system is configured with CUDA
\code{13.0}
and NVSHMEM \code{v3.3.24} and \ToolName{} is compiled with clang \code{20.1.8}.

\paragraph{Multi-node InfiniBand platform}
Scaling experiments on NVIDIA hardware are done on a cluster
equipped with four H100 GPUs and four Mellanox ConnectX-7
MT2910 HCAs per node, connected via NDR 400\,Gb/s InfiniBand.
Communication is handled by NCCL~\code{2.24.3} over
libibverbs from rdma-core~\code{1.14.47.0} and UCX~\code{1.16.0}.
The system runs CUDA \code{12.9} and \ToolName{} is compiled with
gcc \code{13.2.0}.

\paragraph{Multi-node Slingshot platform}
RCCL experiments run on a cluster with four AMD MI250X GPUs per
node connected via an HPE Slingshot interconnect, with
communication handled by RCCL over \code{libfabric}.

\textbf{Tracing modes.}
Tracing experiments compare three \ToolName{} tracing modes
against an uninstrumented baseline where applicable. The modes are described
in \S\ref{sec:design-selective-tracing}: \emph{Aggregate} (thread-local
counters only), \emph{Simple} (timestamped operation events,
excluding high-frequency polling), and \emph{Full} (all
  intercepted calls including individual \code{ibv_post_send} work requests
and \code{ibv_poll_cq} completions).

\subsection{Overhead Study 1: Doorbell Interception Microbenchmark}
\label{sec:evaluation-doorbell}

UAR interposition (\S\ref{sec:doorbell-hijack}) lies on the
latency-critical path of every GPU-initiated RDMA operation.
Its overhead has two contributors:
(1) the UAR forwarding path itself adding an extra write over PCIe
from the CPU polling
thread, and (2) the device-to-host SQ copy, where the polling thread
copies and reads WQE control and data segments.

\paragraph{Per-Operation Latency}
We measure completion latency for four representative GPU-initiated NVSHMEM
operations, each exercising a distinct RDMA verb and WQE type.
All timings are taken by GPU threads using the
\code{globaltimer} register, which provides precise fixed-frequency
nanosecond-resolution timestamps. Each operation is timed
individually inside a persistent
single-threaded kernel with $1000$ warmup and $5000$ measured
iterations.

\begin{itemize}[leftmargin=*]
  \item \textbf{Put+Quiet:} %
    RDMA write followed by a NIC completion wait
    (\code{nvshmem_int_p} then \code{nvshmem_quiet})
  \item \textbf{Get:} a blocking RDMA
    read that returns a remote value with \code{nvshmem_int_g}.
  \item \textbf{Atomic FetchAdd:}
    a blocking RDMA atomic operation that returns the prior value
    (\code{nvshmem_int_atomic_fetch_add}).
  \item \textbf{Ping-Pong:} Round-trip time measurement with
    \code{nvshmem_int_p}
    where ranks write, spin, and write back in order.
\end{itemize}

We test two \ToolName{} configurations against the uninstrumented
baseline: \emph{UAR forward} (doorbell redirect and forwarding only,
no WQE inspection) and \emph{Forward + Read WQEs} (doorbell
forwarding, WQE recovery, parsing, and trace recording).

\begin{figure}[t]
  \centering
  \subfloat[Per-operation completion latency for four NVSHMEM
      operations.  4$\times$A100, single ConnectX-5, GPU-timed via
    \code{globaltimer}.\label{fig:doorbell-latency}]{%
    \includegraphics[width=\linewidth]{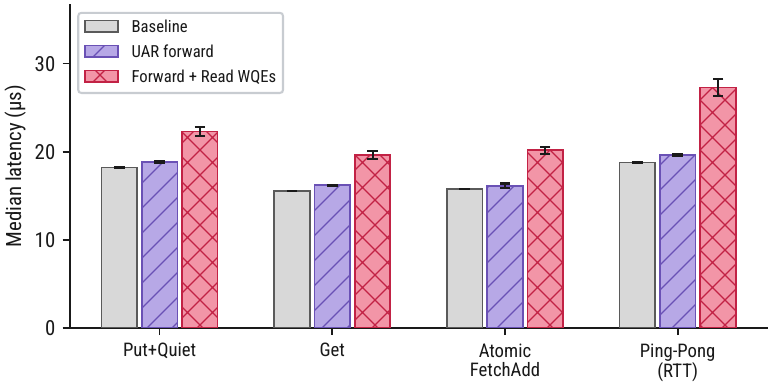}}
  \par\medskip
  \subfloat[Relative overhead vs. baseline for each operation type.
      UAR forwarding adds 2--4\%; forwarding with WQE recovery and
      recording (Forward + Read WQEs) adds 22--46\%.\label{fig:doorbell-overhead}]{%
    \includegraphics[width=\linewidth]{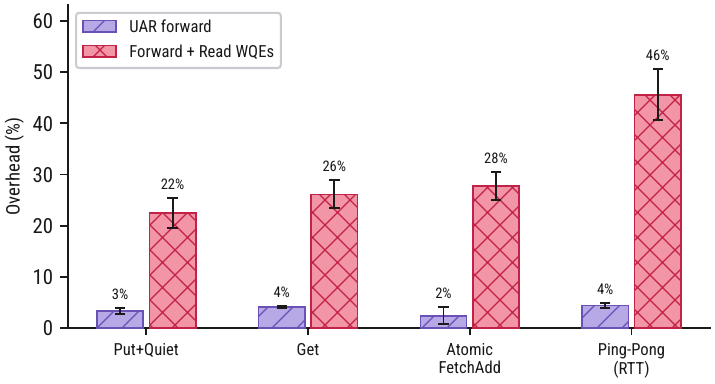}}
  \caption{Per-operation Doorbell Microbenchmark}
\end{figure}

Figure~\ref{fig:doorbell-latency} reports median completion latency.
The baseline latencies are ${\sim}$15--20\,$\mu$s for scalar
operations (\code{put}, \code{get}, \code{atomic})
and higher for ping-pong round trips. Figure~\ref{fig:doorbell-overhead}
shows the relative overhead. UAR forwarding alone adds 2--4\% overhead
across all operation types. This is a fixed per-doorbell overhead from
the polling thread's wake-up latency and the additional PCIe MMIO
write that forwards the doorbell value to the real UAR.
Because GPU doorbell writes themselves are asynchronous, the overhead
does not stall the issuing store; it appears only in operations that
wait on NIC completion (quiet, get, atomic, ping-pong), which is why
all four measured operations are affected.

UAR forwarding combined with reading and parsing the related work
queue entries adds 22--46\% depending on the operation,
even though our batching strategy ensures that synchronized copies
are infrequent relative to doorbell writes.
The additional cost over UAR-only forwarding stems from reading
back WQE control and data segments from GPU-visible memory
over PCIe and recording the trace. Ping-pong exhibits the
highest relative overhead of 46\% as
both the outgoing put and the incoming response traverse the
interposition path, adding to the per-doorbell cost on both
ends of the round trip.

\paragraph{Discussion}

The measured overhead in the above section represents a worst case of
what real deployments experience. Our single-NIC configuration forces
all four GPUs on the system to issue doorbells through one HCA,
which increases the PCIe backpressure and doorbell contention on
polling threads. Realistic deployments distribute doorbells across
multiple HCAs and route a substantial portion of traffic through peer
memory, which is disabled in this worst-case configuration.

\subsection{Overhead Study 2: Multi-Node Scaling}
\label{sec:evaluation-scaling}

While \ToolName{} minimizes cross-rank overhead by design---each
profiler rank writes to a private thread-local ring buffer without
synchronization on the recording path---library interposition, event
collection, and file I/O still introduce overheads at each
instrumented function call.

We run the standard NCCL \code{all_reduce_perf} benchmark on our
multi-node platforms, scaling from 2 to 32 nodes with communication
sizes of 1\,MB and 128\,MB. We report overhead as the percentage
increase in median execution time relative to the uninstrumented
baseline at the same node count.

\begin{figure}[htbp]
  \centering
  \includegraphics[width=\columnwidth]{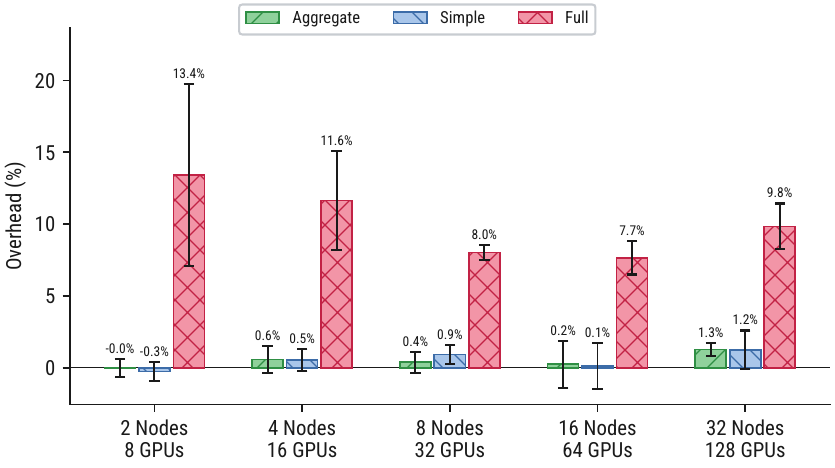}
  \caption{
    NCCL AllReduce overhead on the InfiniBand platform with 1\,MB
    messages. Aggregate and Simple modes stay below 1.5\% at all
    scales, while Full mode incurs higher overhead due to frequent
    event recording by NCCL proxy polling threads.
  }
  \label{fig:nccl-1m}
\end{figure}

\begin{figure}[htbp]
  \centering
  \includegraphics[width=\columnwidth]{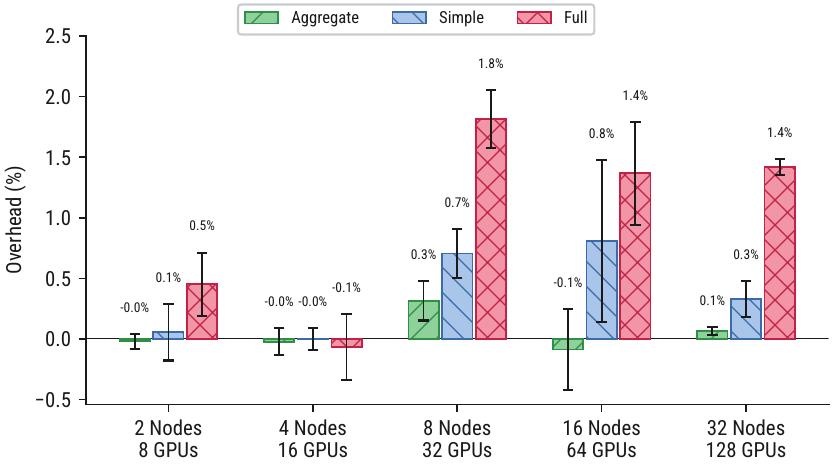}
  \caption{NCCL AllReduce overhead on the InfiniBand platform with
    128\,MB messages. All tracing modes remain below 2\% at every scale
  point.}
  \label{fig:nccl-128m}
\end{figure}

Figure~\ref{fig:nccl-1m} shows the scaling results for 1\,MB messages.
Aggregate and Simple (Default) modes both remain below 1.5\% across
all tested scales, with no upward trend as node count increases from 2 to 32.
Full mode incurs 8--13\% overhead, attributable to recording large
numbers of \code{ibv_poll_cq} completions
and \code{ibv_post_send} and \code{ibv_post_recv} work request data.
Figure~\ref{fig:nccl-128m} shows the results for 128\,MB. At this
message size, all three tracing modes remain within
2\% of baseline at every scale point.
Table~\ref{tab:trace-volume} shows the log sizes and event counts for
both message sizes.
In Full mode, roughly half of the events are \code{ibv_post_send}
work requests (WRs).

\begin{figure}[t]
  \centering
  \includegraphics[width=\columnwidth]{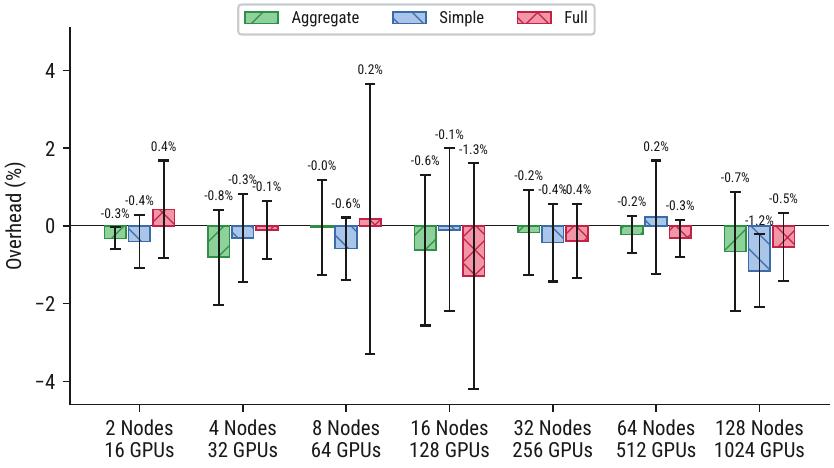}
  \caption{RCCL AllReduce overhead on the Slingshot platform with
    128\,MB messages. All tracing modes stay below 2\% at every
    scale point and within run-to-run variance of the baseline.}
  \label{fig:rccl-128m}
\end{figure}

We additionally run the same AllReduce workload with RCCL over
\code{libfabric} on the Slingshot platform, scaling from 2 to 128
nodes (16 to 1024 GPUs) with 128\,MB messages.
Figure~\ref{fig:rccl-128m} shows results where the overhead of all
three modes stays below $2\%$ at every scale point.
The instrumented call rate on this path is far lower than on the
InfiniBand platform: at the 128\,MB data size, the Profiler collected
16\,MB of data consisting of 17,207 \code{fi_recvmsg} and 4,811
\code{fi_sendmsg} events. Beyond overhead, this experiment validates
the \code{libfabric} instrumentation path and the event collection
pipeline at 1024-GPU scale.

\begin{table}[t]
  \centering
  \caption{Trace size and event counts collected by \ToolName{} on
  32~nodes (128~GPUs) for two message sizes and three tracing modes.}
  \label{tab:trace-volume}
  \begin{tabular}{l l r r}
    \toprule
    \textbf{Buffer Size} & \textbf{Mode} & \textbf{Trace Size} &
    \textbf{Event Count} \\
    \midrule
    1\,MB   & Aggregate & 3.5\,MB  & 50\,K \\
    1\,MB   & Simple    & 1.7\,GB  & 43\,M \\
    1\,MB   & Full      & 13\,GB   & 305\,M \\
    \midrule
    128\,MB & Aggregate & 3.5\,MB  & 42\,K \\
    128\,MB & Simple    & 11\,GB   & 292\,M \\
    128\,MB & Full      & 81\,GB   & 2,000\,M \\
    \bottomrule
  \end{tabular}
\end{table}

\subsection{Case Study 1: Feature Showcase with Distributed BFS}
\label{sec:evaluation-bfs}

GPU-initiated communication can introduce communication and
synchronization stalls that are hidden inside kernel execution and
therefore invisible to many profiling tools. This limitation is
particularly problematic for persistent kernels \cite{cpu-free}, and
tightly coupled kernels that rely on device-side communication to
make joint progress. Irregular applications exhibit this behavior:
communication patterns in graph traversal depend on
the evolving frontier and cannot be inferred from the host-side timeline.
We demonstrate \ToolName{}'s tracing and analysis capabilities on a
distributed Breadth-First Search (BFS) implementation based on
persistent NVSHMEM kernels across 4 GPUs, traced in Simple mode with
doorbell collection enabled.

\begin{figure}[t]
  \centering
  \subfloat[Nsight Systems timeline showing the entire BFS application execution.%
    \label{fig:usecase-2-bfs-nsys}]{%
    \includegraphics[width=\linewidth]{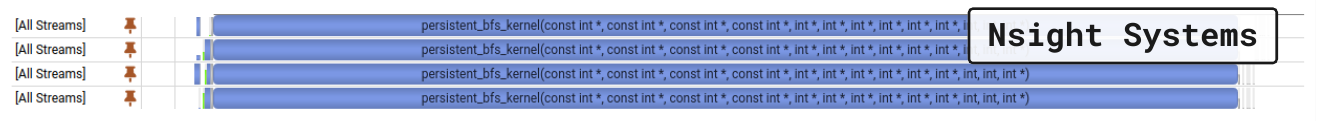}}
  \par\medskip
  \subfloat[\ToolName{} timeline (Simple mode) showing per-kernel
      RDMA doorbell
      traffic and per-GPU bandwidth. Triangle markers denote
      doorbells. Data doorbells (pink) correspond to RDMA \code{READ/WRITE}.
      Control doorbells (brown) correspond to RDMA \code{ATOMIC} and
      barriers. The bottom panel shows the zoomed-in timeline with
      only control doorbells displayed.
    \label{fig:usecase-2-bfs-sigma-1}]{%
    \includegraphics[width=\linewidth]{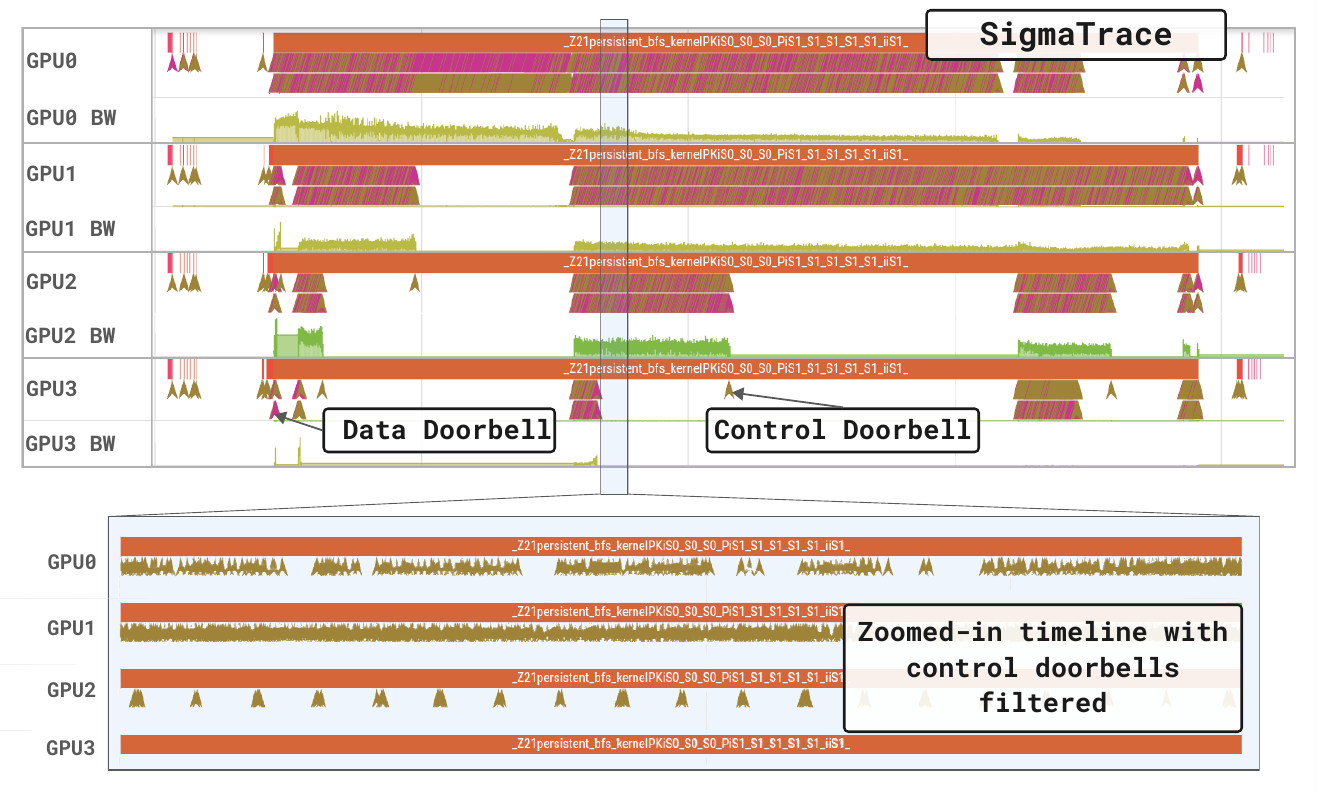}}
  \caption{Comparison of multi-GPU persistent BFS execution as seen
  by Nsight Systems and \ToolName{}.}
  \label{fig:usecase-2-bfs-comparison}
\end{figure}

\paragraph{Workload}

We implement a level-synchronous BFS as a single persistent kernel \cite{atos}.
The graph is stored in Compressed Sparse Row (CSR) format and partitioned across
GPU ranks using 1D vertex partitioning.
At each BFS level, GPU threads expand the local frontier:
edges to local vertices are resolved with device-side atomics,
while edges crossing partition boundaries generate NVSHMEM
atomics. Checks for remote visited vertices generate
additional atomic compare/swap operations, and remote frontier
enqueues are done with memory puts and gets.

We evaluate on {soc-LiveJournal1} \cite{soc-liveJournal}, a directed
social network graph with
4.8\,M vertices and 68\,M edges. The graph exhibits a skewed degree
distribution characteristic
of real-world social networks:
a small fraction of vertices account for a disproportionate share of
outgoing edges.
With 1D partitioning across 4 devices, GPU 0 owns
the highest-degree hub vertices. This creates a structural
communication asymmetry where rank 0 generates substantially more
cross-partition RDMA traffic during early BFS levels as hub vertex
neighborhoods span all remote partitions.

\paragraph{Observation with existing tools}

Figure~\ref{fig:usecase-2-bfs-nsys} shows the Nsight Systems timeline
for a 4~GPU run.
The persistent BFS kernels appear as uninterrupted contiguous blocks
on each GPU,
with almost the same start and end timestamps.
Nsight Systems provides
no further indication of the runtime or communication characteristics
of the kernels: no per-BFS level
timing, no visibility into which GPU generates the most RDMA traffic,
and no information on where
barriers take place or how much overhead they add.

\paragraph{Observation with \ToolName{}}

Figure~\ref{fig:usecase-2-bfs-sigma-1} shows the \ToolName{}
timeline for the same execution. With doorbell collection enabled,
\ToolName{} reveals
the kernels' internal communication structures and provides several
insights into their behavior.

The hub-driven asymmetry is clear as GPU~0 generates large bursts of
RDMA traffic (WRITEs and ATOMICs)
at early levels. GPUs 1 through 3 complete their expansions with
substantially fewer and decreasing number of remote operations.
At deeper BFS levels, the frontier begins to migrate away
from hub vertices. We see this reflected in the RDMA traffic
where GPU~1 traffic equalizes with GPU~0 while GPUs 2 and 3
carry bursts of cross-partition traffic. This level-dependent
behavior is entirely invisible without granular
doorbell timing.

\begin{figure}[t]
  \centering
  \includegraphics[width=\columnwidth]{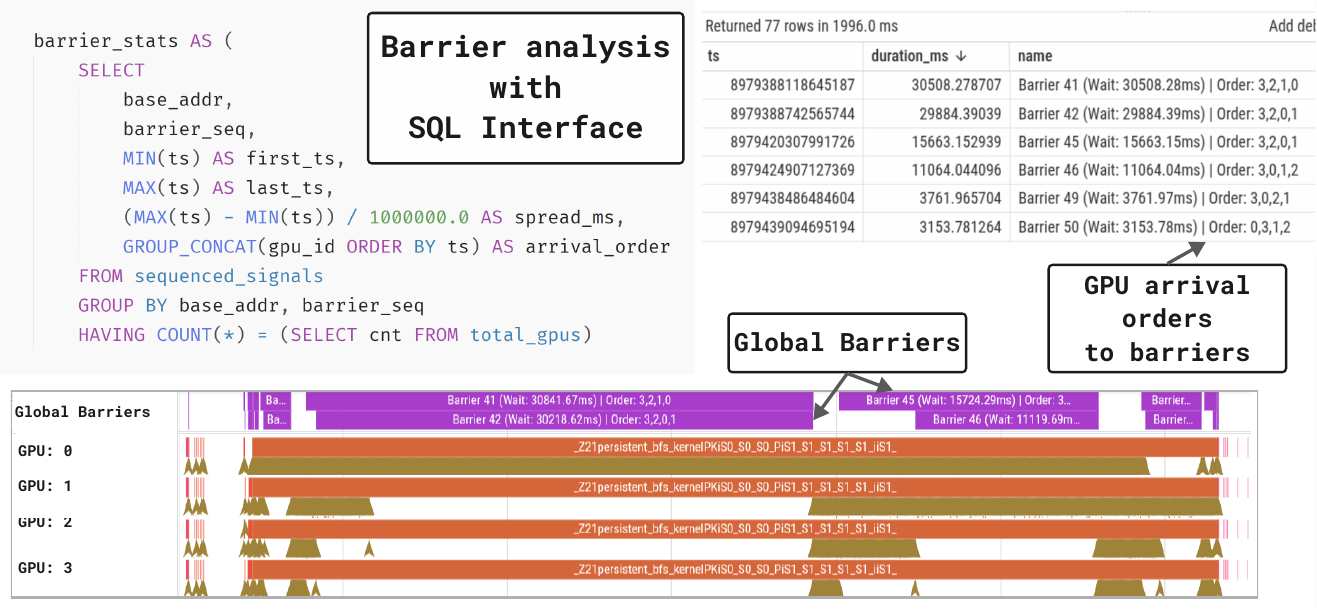}
  \caption{SQL analysis and visualization of global barriers}
  \label{fig:usecase-2-bfs-sigma-sql-1}
\end{figure}

We can filter the doorbells to display only atomic and barrier operations
and zoom into the timeline in Figure~\ref{fig:usecase-2-bfs-sigma-1}.
Since each frontier expansion requires atomic compare/swap, we
observe bursts of atomic doorbell events. The duration
and density of doorbell bursts vary significantly across
GPUs and BFS levels, reflecting the irregular
distribution of cross-partition edges.

\paragraph{Quantitative analysis with SQL}

Figure~\ref{fig:usecase-2-bfs-sigma-sql-1} shows filtered barrier
ranges alongside the
arrival order of the GPUs. The worst-case barrier is at the earliest
levels where GPU~0 dominates
and consequently leaves other peers idle. The barrier spread---the time between
the first and last GPU arrival---reaches 30~seconds at this level.
The pattern begins to invert at deeper levels as the frontier migrates away
from hub vertices.
GPU~1 has the next largest barrier at 15~seconds, followed by GPU~2.

We note that NVSHMEM's block-wide cooperative barrier
(\code{nvshmemx_barrier_all_block}) issues a separate control
doorbell from each calling block, manifesting as overlapping
barrier ranges
in the timeline. We consider the longer-spanning barrier as the
straggler for this application.

\paragraph{Discussion}
\ToolName{}'s doorbell trace and analysis features display
the graph partitioning behavior and inefficiencies reflected in the GPU-initiated RDMA
traffic. The 1D partitioning
concentrates hub vertices on GPU~0, creating early-level imbalance
and large global barriers
at subsequent GPUs. Without per-level and per-GPU doorbell
visibility, the effect of partitioning
strategy on barrier overhead is unobservable.

\paragraph{Overhead}
This experiment, traced in Simple mode with doorbell collection,
incurred an overhead of $11.5\%$ over the baseline, recording
114\,MB of data including 2,718,665 doorbell writes.

\subsection{Case Study 2: Mixture-of-Experts Inference}
\label{sec:evaluation-moe}

Mixture-of-Experts (MoE) inference distributes expert weights across GPUs
and dispatches each token to selected experts. The
resulting network traffic depends on the routing distribution at runtime and
is invisible to existing tools. Using \ToolName{} we can attribute this traffic
to nodes and GPUs, as well as individual QPs.

\begin{figure}[tbp]
  \centering
  \includegraphics[width=0.8\columnwidth]{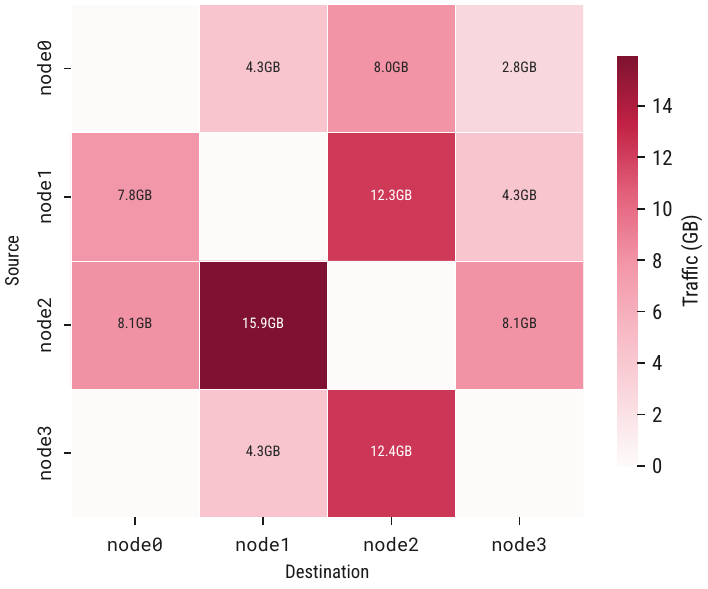}
  \caption{Node-level communication matrix for DeepSeek-V3-0324 served with
  SGLang on 4 nodes / 16 H200 GPUs, driven by 200 ShareGPT requests.}
  \label{fig:eval-moe-matrix}
\end{figure}

\paragraph{Workload} We serve the 671\,B parameter DeepSeek-V3-0324 model
\cite{deepseekai2025deepseekv3technicalreport}, which has 256 routed experts and
one shared expert per MoE layer with top-8 routing. We use SGLang \cite{Sglang}
on 4 nodes with 16 NVIDIA H200 GPUs, configured with data-parallel (DP)
attention and tensor-parallel (TP)-sharded MoE layers,
using host-side NCCL collectives for dispatch and combine. We drive the
server with 200 requests from the ShareGPT dataset and trace in
Aggregate mode.

\paragraph{Observation} Figure~\ref{fig:eval-moe-matrix} shows the
\ToolName{} node-level communication matrix, with bootstrap and
weight-distribution traffic excluded.
The matrix reveals substantial inter-node imbalance: total egress varies by
2.13$\times$ across nodes, and the heaviest pair---node2$\,\rightarrow\,$node1---%
carries 2.15$\times$ the mean per-pair volume. The asymmetry continues
at the cell level, where node2$\,\rightarrow\,$node1 transfers 15.9\,GB while
node1$\,\rightarrow\,$node2 transfers only 12.3\,GB. Node2 is the heaviest node
in both directions.
This indicates uneven token load across the DP ranks: the ranks
hosted on node2 process a larger share of the tokens, generating
proportionally more dispatch and combine traffic to and from their
peers.

\paragraph{Discussion} With the serving stack's scheduler spreading
requests across DP ranks, averaging over many ShareGPT requests
should in principle smooth per-rank token load. The 2$\times$ per-node
imbalance observed here, however, shows that this does not apply to
our configuration. While DP schedulers balance request or token
counts across ranks, token count does not directly imply byte-volume
or NIC-utilization balance: sequence lengths vary per request,
dispatch and combine payloads scale differently, and per-node NIC
placement further shapes the realized network load. \ToolName{} directly
measures the network-side outcome that such balancers ultimately target.

\section{Related Work}
\label{sec:related-work}

Mycroft \cite{Mycroft} and NCCL Inspector
\cite{NVIDIA-Blog-NCCL-Inspector} are NCCL-specific
tools that target activity at host-side proxy threads. Mycroft uses source
code patching while NCCL Inspector
makes use of the NCCL plugin API. Unlike \ToolName{}, neither tool
can observe GPU-initiated operations,
though they gather semantic NCCL data that may be obfuscated within
lower-level traces we collect.
NPKit \cite{NPKit} inserts compile-time macros into \code{*CCL} proxy
threads and GPU kernel source code to emit timed events. It has broader
support with NCCL and MSCCL, but requires recompilation of the target
library with instrumentation macros enabled.

Aegis \cite{Aegis} is a production fault
diagnosis system for AI training infrastructure. It records
RDMA work request metadata and per-collective timing to detect
anomalies such as straggler ranks and network failures. Like Mycroft,
it requires
modifications to NCCL source code and does not observe GPU-initiated paths.
Its focus is fault diagnosis rather than performance profiling.

ucTrace \cite{ucTrace} profiles UCX communication by instrumenting
its transport (UCT)
and protocol (UCP) layers, providing per-operation visibility into
UCX endpoints, tagged
messages, and RDMA operations. While ucTrace covers a broader set of
transports than \code{*CCL}-specific tools, it operates at the UCX
level, and does not see the underlying RDMA verbs or doorbell-level activity.
OSU INAM \cite{osu-inam} profiles MPI applications across the
software stack, correlating MPI-level metrics with fabric-level
counters in near real time. Like ucTrace, it observes host-side
communication only and cannot attribute traffic to GPU-initiated
operations.

Nsight Systems \cite{NVIDIA-Nsight-Systems} provides comprehensive
traces of CUDA APIs, kernel launches, data transfers, collectives, and custom
NVTX annotations. For network observability, it exposes system-level
NIC throughput counters and Mellanox-specific congestion events.
However, these views are aggregated across all processes sharing the
NIC and cannot be attributed to individual GPUs, kernels, QPs, or
remote endpoints. Nsight Systems also cannot observe GPU-side RDMA.
On AMD systems, the ROCm Systems Profiler \cite{AMD-rocprofsys}
samples OS-level NIC interface counters while ROCProfiler
\cite{AMD-rocprof} provides API and kernel traces. Notably, ROCm Systems Profiler
supports device-level RDMA transport counters on AMD's own Pensando NICs, though
neither tool offers per-operation or verbs-level visibility.

Snoopie \cite{Snoopie} is the tool closest in spirit to \ToolName{}
as it tracks GPU-side memory operations using NVBit dynamic binary
instrumentation \cite{NVIDIA-NVbit}. It records every load and store
in a kernel and classifies accesses as local or remote P2P. While
Snoopie's methodology can theoretically be extended to capture writes
to UAR doorbells, instrumenting every memory instruction incurs
overhead that is orders of
magnitude higher than \ToolName{}'s selective doorbell-level interposition.

\section{Limitations}
\label{sec:limitations}

\subsection{Hardware scope}
\label{sec:limitations-hardware}

GPU-initiated RDMA tracing in \ToolName{} is currently only supported on
NVIDIA GPUs with Mellanox NICs.
This covers all GPU-initiated RDMA paths available on NVIDIA platforms.
On AMD GPU systems, rocSHMEM \cite{AMD-rocSHMEM-Docs} implements a
similar GPU-initiated path over Mellanox NICs: its GPU-IB backend
links against the same \code{mlx5dv} and \code{libibverbs} layers and
implements IBGDA identically to NVSHMEM down to the send and
completion queue configuration, memory layout, and WQE sizes. Its newer
GDA backend likewise constructs WQEs and rings doorbells over PCIe from
device code on the same NICs. The NIC-side interception therefore transfers
directly to such systems, but the GPU-side
instrumentation %
would need to be replaced with equivalent ROCm/HIP
driver hooks. We note that the HIP host memory API is one-to-one with CUDA's in
both interface and semantics (\code{hipHostRegister} and
\code{hipHostGetDevicePointer} are analogous to the calls we
intercept). RCCL \cite{AMD-RCCL-Docs} does not currently support
GPU-initiated RDMA and relies on CPU proxy threads, which are
already visible to \ToolName{}'s host-side
instrumentation.

Several other NIC architectures support GPU-initiated RDMA
but are not yet covered by \ToolName{}:
Broadcom Thor 2 \cite{Broadcom_Thor_2}, which uses a different
driver stack, doorbell format, and WQE structure; and
AMD Pensando Pollara 400 \cite{AMD_Pollara_400}, a fully
programmable RDMA NIC with a register interface
unrelated to \code{mlx5dv}. While the GPU-NIC communication model is
fundamentally similar (WQE construction in device memory
followed by a PCIe doorbell write), and the interception mechanism
itself is otherwise hardware-independent, the interposition layer
would require per-NIC adaptations for the doorbell layout and
WQE parsing.

More notably, a GPU-initiated data path for AWS's Elastic Fabric Adapter
(EFA) shipped in NCCL~2.31 as the EFA GDA backend:
CUDA kernels build 64-byte EFA send WQEs and ring the send-queue doorbell
over the SRD transport with no host involvement \cite{efa-dp-direct, aws-ofi-nccl}.
This path is closer to \ToolName{}'s existing target than Thor~2 or Pollara~400---the \code{libfabric} EFA provider hands
the application the send-queue and doorbell BAR addresses through its
\code{FI_EFA_GDA_OPS} extension \cite{libfabric}, which the host
plugin then makes GPU-visible with the \code{cuMemHostRegister}
and \code{cuMemHostGetDevicePointer} calls that \ToolName{} already
intercepts (\S\ref{sec:memory-hijack}). A port would need to hook
\code{libfabric} or \code{efadv} in place of \code{mlx5dv} and decode
the \code{efa_io_tx_wqe} descriptors. We note that in EFA the transport
rings a single 32-bit producer index with no separate doorbell record,
which is simpler than the \code{mlx5dv} workflow.

\subsection{Software scope}
\label{sec:limitations-software}

The driver interception and memory hijack techniques
described in Section~\ref{sec:cuda-interception} and
Section~\ref{sec:memory-hijack} are CUDA-specific. We believe
the methods are sufficiently general to adapt to other
GPU software stacks: the core technique (replacing a
device-visible memory mapping with a monitored proxy page)
is independent of the vendor.  An AMD port would require
replacing \code{cuGetExportTable} interception with
equivalent ROCm driver hooks
and substituting the CUPTI-based GPU event correlation with its
ROCm counterparts (ROCProfiler/rocTracer). Because ROCm is open
source and mirrors the CUDA driver interfaces closely, we expect
this port to be an engineering effort rather than a redesign,
though we have not yet validated this path.

\subsection{Instrumentation scope}
\label{sec:limitations-instrumentation}

We primarily use \code{LD_PRELOAD} to instrument communication
library functions, which is only
compatible with libraries that expose symbols for dynamic linking. While the
overhead is minimal, statically linked libraries and
compiler-inlined function calls cannot be instrumented by
symbol preloading.  The libraries targeted in this work are
designed to be dynamically loaded; however, we found that
\texttt{libcxi} \cite{libcxi-github}, the low-level interface for HPE Cray
Slingshot systems, makes heavy use of \code{static inline}
function definitions, which are invisible to our interposition
layer.

As a result, \textsc{SigmaTrace} cannot currently provide the
same depth of tracing on HPE Cray systems as it does for
\code{libibverbs} and \code{mlx5} direct verbs, though we do provide partial
observability through \code{libfabric}-level instrumentation.
Full doorbell-level coverage and static-compilation support are
porting tasks possible through dynamic instrumentation
(Frida \cite{Frida} or eBPF probes), which can reach inlined
functions at the cost of higher per-call overhead. Combining
\code{LD_PRELOAD} for supported libraries with dynamic probes for
inlined calls is a viable direction for future work.

\section{Conclusion}
\label{sec:conclusion}

We presented \textsc{SigmaTrace}, a library-agnostic RDMA
profiling framework that provides fine-grained observability
into both CPU and GPU-initiated networking. By targeting the
\code{libibverbs} and \code{mlx5dv} direct verbs layers---the common
denominator beneath all InfiniBand and RoCE communication
libraries---\ToolName{} captures per-operation RDMA
behavior without requiring source modifications, recompilation,
or binary instrumentation with $1.3\%$ overhead on NCCL AllReduce at 32 nodes.

Our core contribution is the UAR memory interposition technique that replaces
PCIe MMIO mapped memory pages with CPU-monitored proxy pages,
enabling the first-ever
observability into GPU-initiated RDMA operations at overheads orders
of magnitude below those reported for binary instrumentation
\cite{Snoopie}. We demonstrate that the interposition method
adds 11.5\% end-to-end overhead on a worst-case distributed BFS application
while collecting
previously invisible doorbell-level information that reveals hidden
load imbalance
and computation/communication inefficiencies.
\ToolName{} is publicly available at
\url{https://github.com/ParCoreLab/SigmaTrace}.

\section*{Acknowledgment}
\label{sec:acknowledgment}

This project has received funding from the European Research Council
(ERC) under the European Union's Horizon 2020 research and innovation
programme (grant agreement No 949587).

ChatGPT and Claude were used in preparation of the manuscript.
The tools were used to correct typos and grammar and to polish
wording throughout the text. Additionally, portions of the experiment
drivers and plotting scripts that produced Figures 4--7 and 10 were
generated with these tools. No experimental data, or citations
were generated by an AI system.

% Start balancing in the first column of the final page. Recheck this
% reference index after changes that alter pagination.
\IEEEtriggercmd{\balance}
\IEEEtriggeratref{32}
\bibliographystyle{IEEEtran}
\bibliography{IEEEabrv,references}

\end{document}